\documentstyle[prl,aps]{revtex}
\begin{document}
\title{Dynamical Instabilities 
and Deterministic Chaos
in Ballistic Electron Motion in Semiconductor Superlattices }
\author{Kirill~N.~Alekseev${}^{a,b}$\cite{email1}, 
Gennady~P.~Berman${}^{a,b}$, and David K. Campbell$^{c}$ }
\address{
${}^{a}$Center for Nonlinear Studies, Los Alamos National Laboratory,
Los Alamos, New Mexico 87545\\
${}^{b}$Theory of Nonlinear Processes Laboratory,
Kirensky Institute of Physics, Krasnoyarsk 660036, Russia\\
$^c$Department of Physics,
University of Illinois at Urbana-Champaign,
1110 West Green St.,
Urbana, IL 61801-3080}
\maketitle
\begin{abstract}
We consider the motion of ballistic electrons within a
superlattice miniband under the influence of an
alternating electric field. We show
that the
interaction of electrons with the self-consistent electromagnetic field
generated by the electron current may lead to the transition from regular to
chaotic dynamics.  We estimate the conditions for the experimental
observation of this deterministic chaos and discuss
the similarities of the superlattice system
with the other condensed matter and quantum optical 
systems.
\end{abstract}
\section{Introduction}
Recent progress in the fabrication of semiconductor superlattices has
made
investigations of the physical consequences of ballistic electron
transport in these systems an area of considerable and growing interest. 
Studies of the interaction between ballistic electrons
in superlattices and the electromagnetic field have already
established that nonlinear
phenomena may appear even for small field strength [1]. Theoretical
investigations in this area have examined oscillations of electrons due to
their interaction with constant [2] and alternating [3] electric
fields, self-consistent electron-field oscillations 
[4,5], propagation
of electromagnetic solitons through the superlattice [6], and other
problems. Reviews of the early theoretical investigations of
nonlinear effects in  superlattice - field interactions
have already appeared [7,8].
Recently, oscillations of electrons in superlattices under the influence of
constant electromagnetic field (Bloch oscillations) were 
demonstrated experimentally [9].
\par
It is well known, that for most of nonlinear dynamical systems with more
than  $1.5$ degrees of freedom, transitions from the regular to the chaotic motion
can take place [10]. Among quantum mesoscopic systems, for example,
the possibility that nonlinear effects can lead to instabilities and,
in particular, to dynamical chaos has been widely studied in
Josephson junctions under the presence of an rf-field in a
variety of papers involving
theoretical modeling, numerical simulations, and experiments
(for a review see, e.g. [11]). Several articles have also
studied the possible occurrence of
chaotic dynamics in superlattice-field interactions [12-15].
In particular, chaotic motion of ballistic electrons in a 2D
superlattice in a constant magnetic field was studied within the
classical approximation [12], and recently this approach
has been used to explain the experiments on 
magnetotransport in antidot arrays [13]. 
Other studies suggest that the transition to
dynamical chaos can occur for superlattice electrons
under the influence of a constant magnetic field applied
perpendicular to the axis of superlattice and for either electromagnetic
waves of constant amplitude [14] or for electromagnetic solitons [15].
Chaotic Larmor oscillations of the superlattice electrons [14,15] appear,
when the carriers are initially populated rather close to the top of the
superlattice miniband.
\par
In this paper we discuss another possible manifestation of dynamical
chaos in superlattices: namely, the transition to
chaotic behavior for ballistic electrons moving through a
superlattice and interacting with spontaneously generated self-consistent field
and with the external alternating
electric field. We consider the specifically the case in which
this transition to chaos may occur for electrons with energies
belonging to the bottom of superlattice miniband.

\section{Basic Equations}

Consider the motion of electrons  within the miniband of 1D superlattice under
the the influence of an external alternating electric field
applied along the superlattice axis $z$.  Neglecting
the inter-miniband transitions, the electron dynamics
for weak fields may be described
semiclassically using the effective Hamiltonian
approach [8,16]
\begin{equation}
H =  \epsilon ( {\bf p}) + U({\bf r}, t),\quad {\bf p}=
{\bf P}+{{e}\over{c}}{\bf A}({\bf r},t),
\end{equation}
\begin{equation}
\frac{d{\bf P}}{dt} = - \frac{\partial H}{\partial {\bf r}}, 
\quad {\bf v} \equiv \frac{d{\bf r}}{dt} =  \frac{\partial H}{\partial 
{\bf P}},
\end{equation}
where $\epsilon ({\bf p})$ is the (mini)band energy, ${\bf P}$ is
the canonical quasi-momentum of the electron, $ e $ is the
modulus of the electron charge, ${\bf v}$ is
the electron velocity, and ${\bf A} ({\bf r}, t)$ is
the vector potential due to the electromagnetic field. The
crystal momentum of the electron $\hbar {\bf k}$ is related to
the canonical electron momentum ${\bf P}$ by
\begin{equation}
\hbar {\bf k}= {\bf p}={\bf P} + \frac{e}{c} {\bf A} ({\bf r}, t).
\end{equation}
The potential energy $ U({\bf r},t) $ in (1) is connected with
the alternating electric field applied along $z$ direction
\begin{equation}
E_{ext}(t)= E_{0} \cos ( \Omega t).
\end{equation}
In (4) $ \Omega $ is the 
frequency of the alternating electric field.
The motion of electrons produces a current density
\begin{equation}
j_{z}=-eNv_{z},
\end{equation}
where $N$ is the number of carriers per unit volume. Consequently,
a ``self-consistent'' field, ${\bf A} ({\bf r}, t) $,
is generated and can be described by
the Maxwell equation
\begin{equation}
{\nabla}^{2} A_{z} - \frac{1}{c^2}  \frac{{\partial}^{2} A_{z}}{\partial t^{2}} = - \frac{4 \pi}{c} j_{z}.
\end{equation}
At the same time, according to (1) and (2) this field
affects the electron motion.
\par
We assume a standard dispersion relation for electron motion within
the miniband corresponding to the tight-binding approximation [7,16], so that 
\begin{equation}
\epsilon({\bf p})=\frac{p_{x}^{2}+p_{y}^{2}}{2m^*}+\frac{\Delta}{2} 
\left[ 1-\cos \left( \frac{p_{z}a}{\hbar} \right) \right],
\end{equation}
where $m^*$ is an effective electron mass, $a$ is a superlattice period 
and $\Delta$ is a miniband width.
\par
From (1),(2),(4),(5)-(7) one can derive
a set of equations describing the interaction of electrons
with the self-consistent field and with the alternating electric field
\begin{eqnarray}
{\nabla}^{2} A_{z} -\frac{1}{c^2}  \frac{{\partial}^{2} A_{z}}
{\partial t^{2}} = - \frac{4 \pi}{c} j_{z}, \nonumber \\
j_{z}=-\frac{eNa\Delta}{2\hbar} \sin \left( \frac{a}{\hbar} P_{z} + 
\frac{ea}{\hbar c} A_{z} \right),  \\
\dot{P_{z}}={{\Delta ae}\over{2\hbar c}}{{\partial A_z}\over{\partial z}}
\sin \left( \frac{a}{\hbar} P_{z} + \frac{ea}{\hbar c} A_{z} \right) 
-eE_{0} \cos (\Omega t).  \nonumber
\end{eqnarray}
Consider a sample with a characteristic size shorter than the wavelength 
of the field $A_z$. For typical mesoscopic samples, this leads
to fields in the microwave or far-IR domain. Then in a first approximation
one can neglect the spatial derivatives in (8)
and rewrite this set of equations as one equation
describing a parametrically forced pendulum
\begin{equation}
\ddot{\Phi}+{\omega}_{E}^{2} \sin \left[ \Phi - \frac{{\omega}_{S}}{\Omega} \sin \Omega t \right]=0,
\end{equation}
where we have introduced the following notations :
\begin{eqnarray}
\Phi=\frac{ea}{c\hbar} A_{z} , \quad {\omega}_{S}=\frac{ea}{\hbar} 
E_{0},\quad 
{\omega}_{E}=\left[ \frac{2\pi e^{2} N a^{2} \Delta}{{\hbar}^{2}} 
\right]^{1/2}.
\end{eqnarray}
The frequency ${\omega}_{S}$ characterizes Bloch oscillations when
the self-consistent field is absent and is often called the Stark frequency
[7,8]. As first predicted by Epshtein [4], in the
{\it absence} of the alternating external field (${\omega}_{S}=0 $), a
periodic energy transfer between
the field and the electrons is possible
with the characteristic frequency ${\omega}_{E}$. We note that nonlinear
oscillations of this type in different
semiconductors with nonparabolic dispersion laws  were previously
considered  by Vatova [17].

\section{Some Analogies with Other Condensed Matter
and Quantum Optical Systems and the Transition to Chaos}

Before examining in detail the nonlinear dynamics implied by Eq. (9),
let us consider some instructive analogies with other systems.
Eq. (9) is exactly the same as
the equation describing a Josephson junction subjected to an rf
field in the non-dissipative limit where the feedback mechanism is formed by
an external circuit.
As noted early, it is well-known that in the perturbed Josephson junction
various dynamical instabilities and even dynamical chaos can be observed [11].
Quite recently
Dunlap et al. [18] suggested connecting a superlattice in series to a
capacitor and subjecting the system to an rf field. The circuit
produces the feedback
mechanism, and as a result, the system [18] demonstrates nonlinear properties,
including the possibility of converting ac frequency to dc voltage.
\par
In these two references, the feedback mechanism is  {\it extrinsic}, in
that it is formed by
the external circuit. In contrast, in our present system,
the feedback mechanism is {\it intrinsic} and is
formed through the influence of the
self-consistent field
on the electrons. This situation is analogous to the quantum optical system
which consists of an ensemble of 2-level atoms interacting with
a self-consistent electromagnetic field
and with an external time-periodic field [19-21].
The strength of the external field is characterized by
the Rabi frequency, which is an analog of the Stark
frequency ${\omega}_{S} $ defined in (10).
The self-consistent oscillations of the field have a characteristic frequency
---the so-called ``cooperative frequency'' --  which is an analog of
the Epshtein
frequency ${\omega}_{E} $ (10), and is also proportional to square root of
the number of particles  ${N}^{1/2}$. The miniband width in the
frequency units $\Delta / \hbar$ is
equivalent to the 2-level transition frequency and, pursuing
the analogy to the end, the value
$ e a $ formally corresponds to the dipole moment of the 2-level transition.
The nonlinear dynamics of the system ``2-level atoms plus self-consistent
field plus external field" in some reasonable approximations can also be
described by the equation of forced pendulum [19-21].
As we shall see below, these previous studies of analogous systems
analogies from
the viewpoint of nonlinear dynamics and transition to chaos 
will provide useful
insights into the {\it a priori} complicated superlattice electron dynamics.
\par
We now turn to a detailed description of
the nonlinear dynamics governed by equation (9).
Using an expansion in terms of Bessel functions, one can rewrite equation (9)
as
\begin{equation}
\ddot{\Phi}+{\omega}_{E}^{2} \sum_{n=-\infty}^{\infty} (-1)^{n} J_{n}(G) \sin (\Phi+n \Omega t)=0,
\end{equation}
where $G \equiv {\omega}_{S} / \Omega  $ and $J_{n}(x)$ is the
standard Bessel function. The case when the external field is
absent $({\omega}_{S}=0)$ was first
considered by Epshtein in [4]. It was shown in [4] that if the self-consistent
field is initially present, then the nonlinear energy exchange between
the ballistic electrons and the field is governed by the pendulum
equation. The case of high-frequency perturbation $\Omega \gg {\omega}_{E}$
has been also
considered by Epshtein in [5]. In this case one can neglect all terms with
$n \neq 0 $ in the expansion (11). Thus the
only effect of the high-frequency perturbation is the renormalization of the
frequency  ${\omega}_{E} \rightarrow {\omega}_{E}  \sqrt{J_{0}(G)} $.
In our case, the natural initial conditions for the eq. (9) are those for which both
the self-consistent field and its  vector potential at $ t=0 $ are absent:
$ E_{sc}(0)=A_{sc}(0)=0 $. (Without loss of generality we assume $A_{sc}(0)=const.=0$). Taking into
account that $E=(-1/c) \dot{A} $, we have $\Phi(0)=\dot{\Phi}(0)=0 $.
These initial conditions correspond
to the elliptic stable fixed point of a pendulum without
perturbation $({\omega}_{S}=0
)$. In the remainder of this paper,
we shall consider only these initial conditions.
In the absence of the external field, the self-consistent field is not
generated and can't influence on the electron motion. But for
${\omega}_{S} \neq 0 $,
the self-consistent field can be generated {\it  spontaneously}, due to
instabilities of the motion for some values of the perturbation frequency
$\Omega$. We first consider the case of $G\stackrel{<}{\sim} 1$ and retain in the Bessel
expansion in (11) only the terms up to $\mid n \mid \leq 2$. Then,
we have from (11) after linearization $(\Phi \ll 1)$
\begin{equation}
\ddot{\Phi}+{\omega}_{E}^{2} J_{0}(G) \left[ 1+ \frac{2 J_{2}(G)}{J_{0}(G)}
 \cos 2 \Omega t \right] \Phi = {\omega}_{E}^{2} J_{1}(G) \sin \Omega t.
\end{equation}
From (12) it is evident that at $\Omega \approx {\omega}_{E} $ and
$\Omega \approx (2{\omega}_{E})/l $ (with $l$ an integer),
instabilities exist corresponding to linear and to parametric
resonances. The strongest resonances occur at
$\Omega \approx {\omega}_{E}$
and at $\Omega \approx 2{\omega}_{E} $.
Of course, the growth of the self-consistent field $\dot{\Phi}$ due to
the linear instabilities will eventually saturate at an amplitude
at which the previously neglected nonlinear terms become
significant.
\par
Let us now return to equations (9),(11) and consider the case  $G \gg 1$. 
This
is the well-known problem of nonlinear resonance crossing, which was one
of the starting points in the investigations of the Hamiltonian chaos [22].
In another physical context, analysis of the
same mathematical problem using the method of nonlinear resonance overlap
(the ``Chirikov criterion'') [22,23] showed [24] that at
\begin{equation}
K \approx \frac{{\omega}_{E}^{2}}{\Omega^{2} \sqrt{G}} \gg 1
\end{equation}
the nonlinear dynamics becomes chaotic for the majority of
initial conditions, and the maximal amplitude of the generated
self-consistent field $E^{max}_{sc}$ in the units of frequency is
\begin{equation}
\frac{e a}{\hbar} \mid E_{sc}^{max} \mid = \mid \dot{\Phi}^{max} \mid \sim G \Omega = {\omega}_{S}.
\end{equation}
Therefore, the maximal amplitude of the chaotic field is of the same
order as the amplitude of the external field
$\mid E_{sc}^{max} \mid \sim E_{0} $.
In contrast, at $K<1$ the nonlinear dynamics is regular, and the maximal
amplitude of the generated self-consistent field
is of the order of or less than the width of a single
nonlinear resonance
\begin{equation}
\frac{ea}{\hbar} \mid E_{sc}^{max} \mid = \mid \dot{\Phi}^{max} \mid 
\stackrel{<}{\sim}{\omega}_{E}.
\end{equation}
As one can see, the basic equation (9) can be transformed
to the equation of the periodically forced pendulum 
\begin{equation}
\ddot{\Psi} + {\omega}_{E}^{2} \sin \Psi = g \sin \Omega t,
\end{equation}
by the substitutions $\Psi = \Phi - G \sin \Omega t $ and
$g \equiv {\omega}_{S} \Omega $. 
\par
As we noted above, in the present physical context
we should solve eq. (9) (or (16)) for the
initial conditions $\Phi(0)=\dot{\Phi}(0)=0$. 
Previously, it has been shown ([20]), for the quantum optical analog of the
model described by  eq. (16), that at 
$\Omega / {\omega}_{E} \stackrel{<}{\sim}1$
dynamical chaos is possible even when at $t=0$ the self-consistent field is
absent. This situation can be realized if the dimensionless perturbation
parameter $g/ {\omega}^2_{E}$ is larger than some critical value
of order one. 
\par
So, for slow external perturbation $(\Omega / {\omega}_{E}
\stackrel{<}{\sim} 1)$
the self-consistent field
can be generated spontaneously and be chaotic. It should be noticed that
in this case the amplitude of the self-consistent field is of the same
order as or larger than the field amplitude for the regular motion. 
\par
The dynamical behavior of the system (9) is illustrated in Figs. 1-4.
The chaotic time-dependence of the spontaneously generated
self-consistent field is shown in Fig. 1a for $\Omega/\omega_E =0.5$.
In contrast, for the same amplitude of the external field, but for
$\Omega / {\omega}_{E}\stackrel{>}{\sim} 1$, the dynamics of the self-consistent field
is regular (Fig. 1b). It is seen from comparison of Figs. 1a and 1b that
the amplitude of the chaotically generated self-consistent field is 
several times larger than the field amplitude of the regular motion.
Fig. 2 demonstrates the regions of regular motion and of strong chaotic
dynamics in the plane of dimensionless parameters $\Omega/\omega_E$
and $\omega_S/\omega_E$, and at fixed initial conditions:
$\Phi(0)=\dot{\Phi}(0)=0$. Figs. 3 and 4 illustrate
the modification of the nonlinear oscillations of the self-consistent field
under the variation of the amplitudes of the external field but at
fixed frequency $\Omega/\omega_E$. Fig. 3 shows the nonlinear dynamics
at rather slow frequency, $\Omega/\omega_E=0.1$. In this case, the
chaotic dynamics exhibits intermittent behavior (see Fig. 3a). In the
region of high-frequency external field ($\Omega/\omega_E\stackrel{>}{\sim} 1$), the
transition from regular to chaotic dynamics is rather sensitive to variations
of the parameters (see Fig. 2). The chaotic dynamics in this case
also reveals the character of intermittency, which is demonstrated in Fig. 4a. 

\section{ Conclusion}

Our analysis establishes that the motion of ballistic electrons
through a semiconductor superlattice can, when one takes account
of the generation of a self-consistent field,
demonstrate both linear and nonlinear instabilities and
deterministic chaos. We have argued that this behavior is quite
analogous to the quantum optical model [20]
describing nonlinear dynamics of 2-level atoms interacting with
the self-consistent and with the external time-periodic fields.
\par
It is important to examine the condition of validity of our
approach. From [7,8], we conclude that this condition requires that all
characteristic frequencies are much less then the miniband
width.  In our case, this condition takes the explicit form
\[
max \left \{\Omega, {\omega}_{E},{\omega}_{S} \right \} \ll \Delta / 
\hbar. \nonumber
\]
For $\Delta \sim 10^{-2} {\rm eV}$,
$N \sim 10^{14}{\rm {cm}^{-3}} $, $ a \sim 10^{-6} {\rm cm}$ and
 $eaE_{0} \sim 1 {\rm meV} $ [3,7], the
Stark frequency $(\omega_{S})$ is much less than $\Delta / \hbar $ but
comparable to the collective Epshtein
frequency $ \omega_{E} $ (10) and thus belongs to the terahertz region. In
this case, the criterion for chaotic dynamics may be satisfied.
So, one can see that this first general condition for a transition to 
chaos in the case
of ballistic electrons in superlattices are close to those
needed for the observation of Bloch oscillations and related phenomena in
superlattices [2,3,7-9,18]. 
For applications to realistic experimental systems, it is essential to
understand and model the role of {\it dissipation} caused by the
collisions of the ballistic electrons with impurities and phonons. In our
present considerations, the neglect of these effects means that our
model nonlinear dynamics actually represents Hamiltonian chaos,
whereas in realistic systems, the dynamics would likely be dominated
by dissipative effects and consequently attractors. Nonetheless, for
parameters similar to those discussed above,
we expect dissipative chaos and resulting
strange attractors to emerge. A more detailed discussion of these
dissipative effects, of analytic estimates of the boundary of the
region of chaos, and of potential applications to experiments
will be presented elsewhere [25].
\section* {Acknowledgments}
We acknowledge fruitful discussions with Mark Sherwin and James Bayfield.
KNA and GPB thank Don Cohen of The Center for Nonlinear Studies, Los
 Alamos National Laboratory, for the hospitality. This work was partially supported by the
Grant 94-02-04410 of the Russian Fund for Basic Research and by
the Linkage Grant 93-1602 from the NATO Special Programme Panel on
Nanotechnology.

\section* {Figure Captions}
\begin{enumerate}
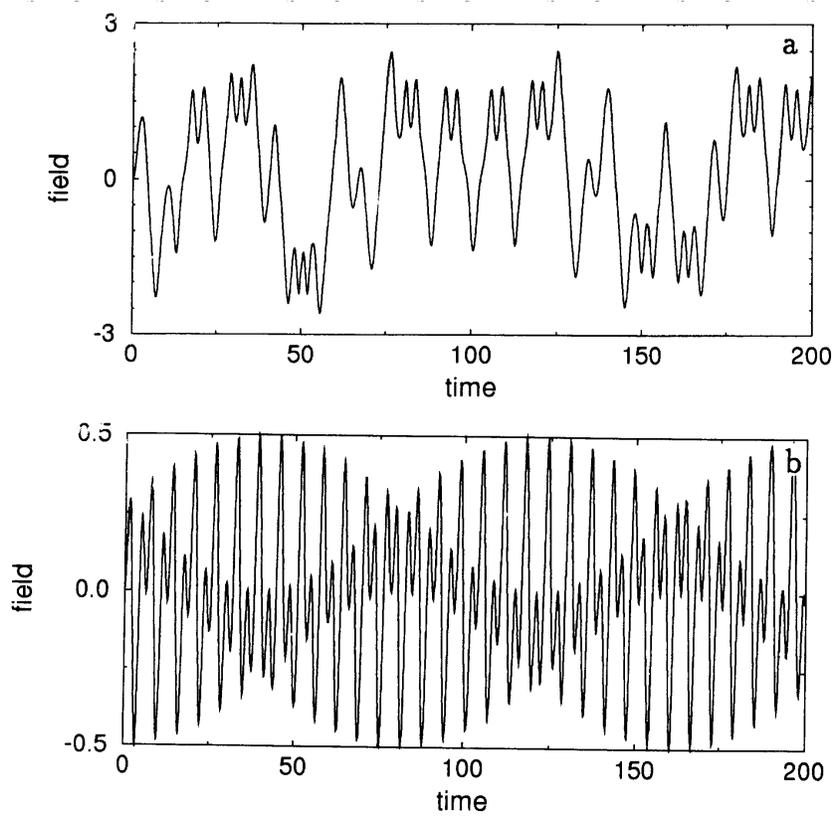
\item [Fig. 1] The dependence of the dimensionless amplitude of the
self-consistent field $ {\cal E}\equiv ea E_{sc}/\hbar \omega_{E}$ on the
dimensionless time $\tau=\omega_E t$: (a) 
chaotic dynamics for $\omega_{S} / \omega_{E} = 0.8 $, and
$\Omega / \omega_{E} = 0.5 $;
(b) regular dynamics for $\omega_{S} / \omega_{E} = 0.8$, and
$\Omega / \omega_{E} = 2 $.
For both cases, the initial conditions are  $ \Phi (0) =
\dot{\Phi}(0)=0$.

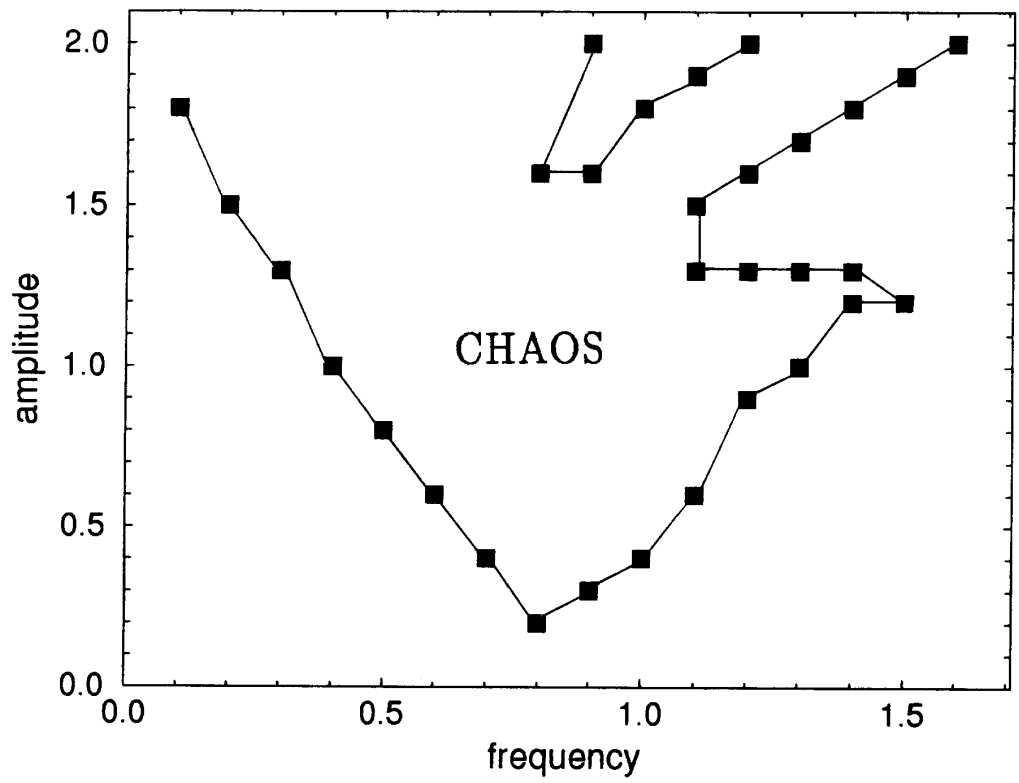
\item [Fig. 2] The regions with strong chaotic and with regular dynamics in
the space of the dimensionless parameters: the frequency of
the external field ($\Omega/\omega_E$), and the amplitude of the
external field ($\omega_S/\omega_E$). The initial conditions are
$ \Phi (0) = \dot{\Phi}(0)=0$.

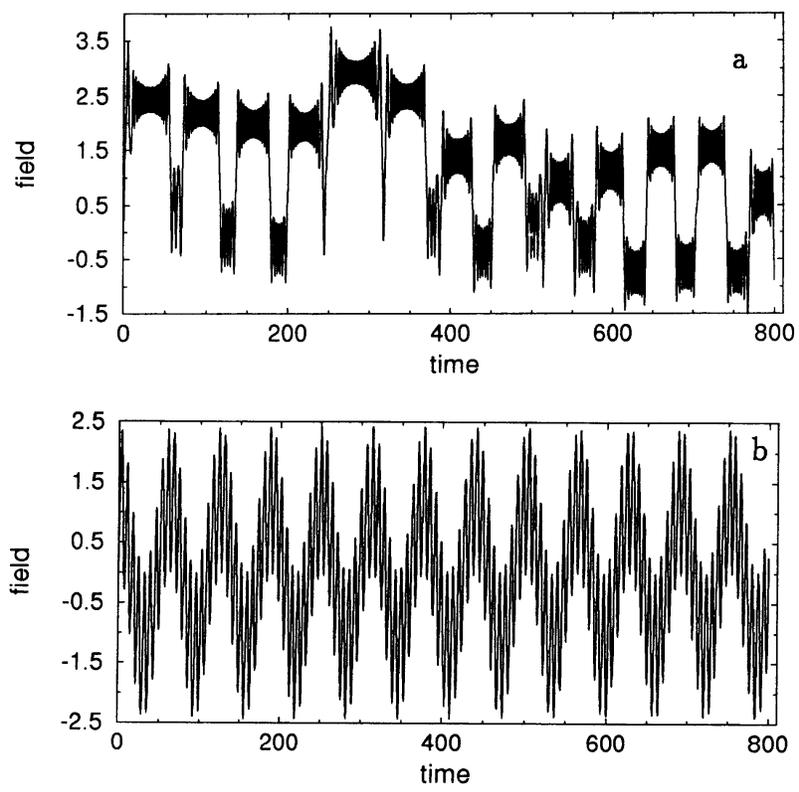
\item [Fig. 3] The same as in Fig. 1 but for the parameters: (a) chaotic
dynamics for $\omega_{S} / \omega_{E} = 1.8$, and $\Omega / \omega_{E} = 0.1 $;
(b) regular dynamics for $\omega_{S} / \omega_{E} = 1.2$, and
$\Omega / \omega_{E} = 0.1 $.

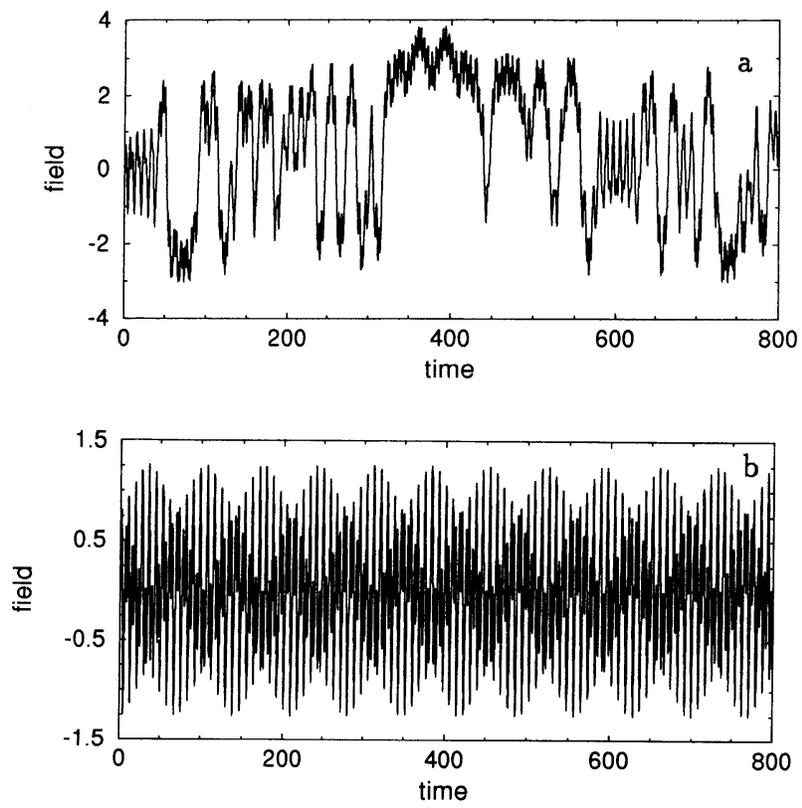
\item [Fig. 4] The same as in Fig. 1 but for the parameters:
(a) chaotic dynamics for $\omega_{S} / \omega_{E} = 1.2$, and
$\Omega / \omega_{E} = 1.5 $; (b) regular dynamics for
$\omega_{S} / \omega_{E} = 1.5$, and 
$\Omega / \omega_{E} = 1.5 $.
 
\end{enumerate}
\end{document}